\documentstyle[11pt,aaspp4]{article}

\newcommand{\etal}{~et~al.}

\newcommand{\wisk}[1]{\ifmmode{#1}\else{$#1$}\fi}

\def\gap{\;_\sim^>\;}
\def\lap{\;_\sim^<\;}

\begin{document}
\vspace{-2.0truecm}
\begin{flushright}
KSUPT-98/4, KUNS-1532 \hspace{0.5truecm} November 1998
\end{flushright}
\vspace{-0.5truecm}
\title{CMB Anisotropy Constraints on Open and Flat-$\Lambda$ CDM Cosmogonies 
       from UCSB South Pole, ARGO, MAX, White Dish, and SuZIE Data}
\author{
  Bharat~Ratra\altaffilmark{1},  
  Rados{\l}aw~Stompor\altaffilmark{2,3},
  Ken~Ganga\altaffilmark{4},
  Gra{\c c}a~Rocha\altaffilmark{3},  
  Naoshi~Sugiyama\altaffilmark{5},
  and
  Krzysztof~M.~G\'orski\altaffilmark{6,7}
  }
\altaffiltext{1}{Department of Physics, Kansas State University,
                 Manhattan, KS 66506.}
\altaffiltext{2}{Institute of Astronomy, University of Cambridge,
                 Madingley Road, Cambridge, CB3 0HA, UK.}
\altaffiltext{3}{Copernicus Astronomical Center, Bartycka 18, 
                 00-716 Warszawa, Poland.}
\altaffiltext{4}{IPAC, MS 100--22, California Institute of Technology,
                 Pasadena, CA 91125.}
\altaffiltext{5}{Department of Physics, Kyoto University,
                 Kitashirakawa-Oiwakecho, Sakyo-ku, Kyoto 606-8502, Japan.}
\altaffiltext{6}{Theoretical Astrophysics Center, Juliane Maries Vej 30,
                 2100 Copenhagen \O, Denmark.}
\altaffiltext{7}{Warsaw University Observatory, Aleje Ujazdowskie 4, 
                 00-478 Warszawa, Poland.}
\begin{abstract}
  We use combinations of ten small-scale cosmic microwave background  
  anisotropy data sets from the UCSB South Pole 1994, ARGO, MAX 4 
  and 5, White Dish and SuZIE experiments to constrain cosmogonies. 
  We consider open and spatially-flat-$\Lambda$ cold dark matter 
  cosmogonies, with nonrelativistic-mass 
  density parameter $\Omega_0$ in the range 0.1--1, baryonic-mass density 
  parameter $\Omega_B$ in the range (0.005--0.029)$h^{-2}$, and age of the 
  universe $t_0$ in the range (10--20) Gyr. 

  Marginalizing over all parameters but $\Omega_0$, the combined
  data favors an $\Omega_0\simeq$ 1 (1) open (flat-$\Lambda$) model. Excluding 
  the smallest angular scale SuZIE data, an $\Omega_0\simeq$ 0.3 (1) open 
  (flat-$\Lambda$) model is favored. Considering only 
  multi-frequency data with error bars consistent with sample variance 
  and noise considerations, i.e., the South Pole 1994 Ka band, the 
  MAX 4 $\iota$ Draconis, and the MAX 5 HR5127 data, an $\Omega_0\simeq$ 
  0.1 (1) open (flat-$\Lambda$) model is favored. For both open 
  and flat-$\Lambda$ models and for all three combinations of data sets, 
  after marginalizing over all the other parameters, a 
  lower $\Omega_B h^2 (\sim 0.005)$ or younger ($t_0 \sim 10$ Gyr) universe is 
  favored. However, the data do not rule out other values of $\Omega_0$ in 
  the flat-$\Lambda$ model and other values of $\Omega_B h^2$ in both models. 
  At 2 $\sigma$ confidence, model normalizations deduced from the 
  small-scale data are consistent with those derived from the DMR data.
  We emphasize that since we consider only a small number of data sets,
  these results are tentative.
\end{abstract}

\section{Introduction}

Six years ago the $COBE$-DMR experiment detected anisotropy in the cosmic 
microwave background (CMB) on angular scales $\sim 10^\circ$ (Smoot 
et al. 1992; Wright et al. 1992; Bennett et al. 1996; G\'orski et al. 1996). 
Since then a number of experiments have measured the anisotropy on smaller 
angular scales, down to arcminutes (Ganga et al. 1994; 
Guti\'errez et al. 1997; Femen\'{\i}a et al. 1998; Netterfield et al. 1997; 
Gundersen et al. 1995; Tucker et al. 1997; de Oliveira-Costa et al. 1998; 
Platt et al. 1997; Masi et al. 1996; 
Lim et al. 1996; Cheng et al. 1997; Griffin et al. 1999; Baker et al. 1998; 
Leitch et al. 1998; Church et al. 1997; Subrahmanyan et al. 1998).

These observations are becoming an increasingly powerful tool 
for testing cosmogonies and constraining cosmological parameters
such as $\Omega_0$, $h$, and $\Omega_B$ in these models\footnote{
Here $\Omega_0$ is the nonrelativistic-mass density parameter, $h$ is the 
Hubble parameter in units of $100\ {\rm km}\ {\rm s}^{-1}\ {\rm Mpc}^{-1}$, 
and $\Omega_B$ is the baryonic-mass density parameter.}.
While definitive results will probably have to await new data 
acquired at a variety of frequencies to constrain or estimate possible non-CMB
anisotropy foreground emission, it is of interest to explore what 
constraints current data place on cosmological model parameters. Given the 
error bars associated with current measurements, interesting 
constraints on cosmological model parameters require 
simultaneous use of many data sets in this exploration.\footnote{
It is important to use as many data sets as possible, and to not rely on just
a few, since any data set could be biased by an as-yet undiscovered systematic
effect.}
Two distinct 
techniques have been used to combine results from different data sets.

In the first approach, a goodness-of-fit (``$\chi^2$") comparison of 
CMB anisotropy predictions, which depend on model parameters, and observational 
results is used to constrain cosmological model parameters (e.g., Ganga, Ratra, 
\& Sugiyama 1996; Hancock et al. 1998; Lineweaver et al. 1997;
Lineweaver \& Barbosa 1998; Baker et al. 1998). This comparison does not
make use of the complete data from each experiment. Rather, it typically
uses a single number (amplitude of a predefined CMB spectrum) with error bars, 
derived from the rms
anisotropy measured by the experiment. As a consequence it is an easily used 
technique that allows for a rapid exploration of cosmological parameter 
space. On the other hand it does have a number of significant drawbacks, 
some of which are discussed, most recently, by Bond, Jaffe, \& Knox (1998) 
and by Tegmark (1998). In particular,
the amplitude (and error bars) used to represent the data is model 
dependent --- this is typically a $\sim 10\%$ effect for current data 
sets with good detections (Ganga et al. 1997a, hereafter GRGS; Ganga et al.
1998; Ratra et al. 1999, hereafter R99). This is not accounted for in the 
$\chi^2$ analyses, since they use an amplitude (and error bars) 
extracted from the data on the basis of an assumed flat bandpower spectrum
or gaussian autocorrelation function. More importantly, these  
amplitudes and error bars are derived from quite nongaussian posterior
probability density distribution functions. As a result the observational error 
bars are quite asymmetric and symmetrizing (``gaussianizing") them in 
different ways results in different $\chi^2$ values (Ganga et al. 1996).
Note that gaussianization is mandatory, not optional, for the $\chi^2$
technique. Furthermore, the $\chi^2$ technique, as currently applied, 
can not account for observational upper limits. As discussed below, at least 
one upper limit (SuZIE) significantly constrains cosmological parameter space. 
In addition, when assigning confidence limits on cosmological parameters
to the constant $\chi^2$ bounds, proponents of the
$\chi^2$ technique assume that the resulting probability distribution is 
a gaussian function of the cosmological parameters, which is not true.
Given these drawbacks
it is clear that one should not rely solely on the $\chi^2$ technique for
quantitative constraints on cosmological parameters. We note, however, that
the technique does qualitatively establish that the CMB anisotropy spectrum 
has more power on smaller angular scales (Ganga et al. 1996), consistent with
recent Saskatoon observations (Netterfield et al. 1997).  

The second and more correct approach, a joint maximum likelihood analysis of 
all the data sets using realistic model anisotropy spectra, is very time
consuming. This approach has been used to constrain cosmological parameters 
on the basis of a single data set (Bunn \& Sugiyama 1995; G\'orski et al. 
1995, 1998; Stompor, G\'orski, \& Banday 1995; Yamamoto \& Bunn 1996; GRGS;
Ganga et al. 1997b, 1998; Bond \& Jaffe 1997; Stompor 1997; Ratra et al. 
1998; R99), and a combination of three data sets (Bond \& Jaffe 1997). 
Such analyses use the full information in each data set, rather
than the single amplitude with error bars used in the $\chi^2$ analyses.

In this paper we combine results from earlier analyses of the Gundersen
et al. (1995) UCSB South Pole 1994 data, the Church et al. (1997) SuZIE data, 
the MAX 4 and 5 data (Tanaka et al. 1996; Lim et al. 1996), the Tucker et
al. (1993) White Dish data, and the de Bernardis et al. (1994) ARGO Hercules
data.\footnote{
We use this set of data as a test case to develop the method, since these
are the results we currently have access to.}
 These analyses made use of theoretically-predicted CMB anisotropy 
spectra in open and flat-$\Lambda$ cold dark matter (CDM) models (GRGS; Ganga 
et al. 1997b, 1998;
Ratra et al. 1998; R99). Since these data sets were acquired from regions
that are well separated in space, the likelihoods of the individual data 
sets are independent and can thus be multiplied together to construct the 
likelihood of the combined
data set. This combined likelihood is then used to derive constraints
on cosmological model parameters.

Our analysis here is complementary to that of Bond \& Jaffe (1997). They 
consider the DMR, UCSB South Pole 1994, and Saskatoon data sets and develop a 
method to constrain cosmological parameters, while we focus here on a larger 
number of smaller angular scale data sets. To cut down on complexity, we do 
not consider the DMR data in this paper. Different ways 
of dealing with the quadrupole moment in the DMR analysis lead to significantly
different constraints on cosmological parameters (G\'orski et al. 1998), so a 
proper treatment would require consideration of at least two different sets
of DMR results, leading to a significant increase in the number of data set
combinations (since we consider various combinations of the small-scale 
data sets) and a proliferation of figures. We also want to first focus on the 
constraints from smaller angular scale data, and, in particular, compare 
results from different combinations of the small-scale data sets.
While we consider fewer model parameters than did 
Bond \& Jaffe --- for instance, we do not allow for tilt, gravity
waves, or hot dark matter --- we explore, in a systematic way, a much broader
range for the model parameters considered. Also, unlike Bond \& Jaffe, 
we perform a likelihood analysis of the complete data, i.e., we do not use
data compression to speed up the computation and minimize memory requirements.

In $\S$2 we describe the models and cosmological parameter space we 
consider. See R99 for further details. In $\S$3 we discuss
the various combinations of data sets we consider.
In $\S$4 we summarize the computational techniques we use. See GRGS for a 
more detailed description. 
Results are presented and discussed in $\S$5 and we conclude in $\S$6.

\section{Cosmogonical Models}

Current data are most easily accommodated in a low-density cosmogony. 
These data include:
\begin{itemize}
\item Dynamical estimates of the mass clustered on scales $\lap$ 10
$h^{-1}$ Mpc, which suggest a low $\Omega_0$. 
For example, virial analysis of X-ray cluster data 
indicates $0.07 \leq \Omega_0 \leq 0.31$ at 2 $\sigma$ (Carlberg et al. 
1997a). Some of the mass might not cluster on these small scales, so these 
estimates might be biased low. However, recent dynamical estimates of the mass 
clustered on large scales $\gap$ 10 $h^{-1}$ Mpc indicate $0.1 \lap \Omega_0
\lap 0.6$ at $\sim 2$ $\sigma$ (e.g., Willick et al. 1997; Borgani et al. 1997,
also see Small et al. 1998).
\item Low estimates of $\Omega_0$ from measurements of the baryonic-mass
fraction of the rich clusters, standard nucleosynthesis, and the observed 
light-element abundances (e.g., Evrard 1997; Ettori, Fabian, \& White 1997).
\item The redshift $z \sim 0$ masses and abundances of galaxy clusters,
which indicate $0.25 \lap \Omega_0 \lap 0.5$ in DMR-normalized CDM cosmogonies 
(e.g., Cole et al. 1997).
\item The shape of the observed galaxy fluctuation power spectrum (e.g.,
Maddox, Efstathiou, \& Sutherland 1996).
\item The lack of large evolution in the galaxy cluster luminosity function,
to $z \sim 0.5$, which indicates $0.2 \lap \Omega_0 \lap 0.6$ at 1.6 $\sigma$ 
in CDM models (e.g., Carlberg et al. 1997b; Ebeling et al. 1997; Mushotzky \& 
Scharf 1997; Bahcall, Fan, \& Cen 1997; Eke et al. 1998, also see Voit \&
Donahue 1998).
\item Indications of high-$z$ structure formation, e.g.,
massive clusters at $z \sim 0.5-1$ (e.g., Deltorn et al. 1997; Donahue et al. 
1998), similarity between the giant elliptical luminosity function
at $z \sim 1$ and the present (e.g., Gardner et al. 1997; Small, Sargent, 
\& Hamilton 1997, also see van Dokkum et al. 1998),
massive galactic disks at $z \sim 1$ (Vogt et al. 1996), 
and galaxy groups at $z > 2$ (e.g., Francis, Woodgate, \& Danks 1997).
\end{itemize}

The simplest low-density CDM cosmogonies have either flat spatial 
hypersurfaces and a cosmological constant $\Lambda$, or open spatial
hypersurfaces and no $\Lambda$. Both these low density models are consistent 
with the data mentioned above. For recent discussions see Park et al. (1998), 
Peebles (1998), Retzlaff et al. (1998), Col\'{\i}n et al. (1998), Croft et al. (1998), and Governato et al. (1998). There is 
additional data which favors either the flat-$\Lambda$ or the open model.

The data which favors the open model include:
\begin{itemize}
\item Measurements of the Hubble parameter which suggest $h = 0.65 \pm 0.1$ 
at 2 $\sigma$ (e.g., Giovanelli et al. 1997; Hjorth \& Tanvir 1997; Falco
et al. 1997; Della Valle et al. 1998), and measurements of age of the universe 
which indicate $t_0 = 12 \pm 2.5$ Gyr at 2 $\sigma$ (e.g., Feast \& Catchpole 
1997; Reid 1997; Gratton et al. 1997; Chaboyer et al. 1998). The resulting 
central $H_0 t_0$ value is
consistent with an open model with $\Omega_0 \approx 0.35$ and a flat-$\Lambda$
model with $\Omega_0 \approx 0.6$, somewhat larger than what is favored by 
the data above. At 2 $\sigma$ there is no significant constraint on
$\Omega_0$ in the open model, but $\Omega_0 \gap 0.25$ is required in the
flat-$\Lambda$ model.
\item Analyses of the rate of gravitational lensing of quasars and radio
sources by foreground galaxies requires $\Omega_0 \geq 0.38$ at 2 $\sigma$ 
in the flat-$\Lambda$ model but only weakly constrains the open case (e.g., 
Bloomfield Torres \& Waga 1996; Kochanek 1996; Falco, Kochanek, \& Mu\~noz 
1998; Jain et al. 1998).\footnote{
It has recently been suggested that systematic uncertainties currently
preclude a strong constraint on $\Lambda$ from the gravitational lensing
of quasars (Cheng \& Krauss 1998). The constraint using only radio data 
is still quite restrictive, $\Omega_0 \geq 0.27$ at 2 $\sigma$ (Falco
et al. 1998).}
\item The predicted number of large arcs formed by strong gravitational
lensing by clusters in the open model is more consistent with what is 
observed (Bartelmann et al. 1998).
\item When normalized to the DMR observations, the flat-$\Lambda$ CDM model 
with a scale-invariant spectrum has excessive intermediate- and small-scale
power and hence requires mild antibiasing, which is not easily reconciled 
with the observations (e.g., Stompor et al. 1995; Liddle et al. 1996; 
Cole et al. 1997).
\end{itemize}

In passing, we note that most of these observations can probably be 
reconciled with a 
time-variable cosmological ``constant" dominated spatially-flat model 
(e.g., Peebles \& Ratra 1988; Sugiyama \& Sato 1992; Ratra \& Quillen 1992;
Coble, Dodelson, \& Frieman 1997; Ferreira \& Joyce 1997; Caldwell, Dave, \&\ 
Steinhardt 1998; Viana \& Liddle 1998; Frieman \& Waga 1998; Anderson \&
Carroll 1998; \"Ozer \& Taha 1998; Huterer \& Turner 1998; Starobinsky 1998). 

On the other hand, recent applications of the apparent magnitude versus 
redshift test using Type Ia supernovae favor the flat-$\Lambda$ model (e.g., 
Riess et al. 1998; Perlmutter et al. 1999).

We emphasize that some, if not most, of the above constraints are very 
tentative. They should be viewed as indicative as what may soon be possible,
and are certainly not definitive. 

In the analyses in this paper we focus on a spatially open CDM model 
and a spatially flat CDM model with a $\Lambda$.
These models have gaussian, adiabatic primordial energy-density power
spectra. The flat-$\Lambda$ model CMB anisotropy computations use a 
scale-invariant energy-density perturbation power spectrum (Harrison 1970;
Peebles \& Yu 1970; Zel'dovich 1972), as predicted in the simplest 
spatially-flat inflation models (Guth 1981; Kazanas 1980; Sato 1981a, 1981b).
The open model computations use the energy-density power spectrum 
(Ratra \& Peebles 1994, 1995; Bucher, Goldhaber, \& Turok 1995; Yamamoto,
Sasaki, \& Tanaka 1995) predicted in the simplest open-bubble 
inflation models (Gott 1982; Guth \& Weinberg 1983). 

To make the problem tractable, in each model (open and flat-$\Lambda$)
we consider anisotropy spectra parameterized by (i) the quadrupole-moment
amplitude $Q_{\rm rms-PS}$, (ii) $\Omega_0$, (iii) $\Omega_B h^2$, and 
(iv) $t_0$. While it is of interest to also consider other cosmological 
parameters, current data does not justify the effort needed to explore
a larger dimensional parameter space. In particular, in our analyses here
we ignore the effects of tilt, primordial gravity waves, and reionization. 
These effects are unlikely to be very significant in viable open models although
they could help reconcile some flat-$\Lambda$ model predictions with the 
observations. More specifically, while we do not use goodness-of-fit 
statistics in this paper to check whether the favored cosmological 
parameter values derived here are a good fit to the data, $\chi^2$ analyses
have qualitatively shown that some of the models in the four-dimensional 
parameter space 
we study in this paper are indeed a good fit to the data, possibly better 
than should be expected (Ganga et al. 1996; Lineweaver \& Barbosa 1998;
Tegmark 1998). That is, current data does not require consideration of
a larger dimensional parameter space. In addition, we note that constraints
on model parameters from each of the individual data sets used in this paper
are largely mutually consistent. 

The computation of the anisotropy spectra is described in Stompor (1994) 
and Sugiyama (1995). We have evaluated the spectra for a range of
$\Omega_0$ spanning the interval 0.1 to 1 in steps of 0.1, for a range
of $\Omega_B h^2$ spanning the interval 0.005 to 0.029 in steps of 0.004, and
for a range of $t_0$ spanning the interval 10 to 20 Gyr in steps of 2 Gyr.
See R99 for further details. Figure 1 shows examples of the model CMB 
anisotropy spectra used in our analyses. Other examples are shown in Figure 2 
of R99.

\section{CMB Anisotropy Data Sets}

We consider various combinations of ten different data sets. The 
data sets we use are the UCSB South Pole 1994 Ka and Q band observations,
hereafter SP94Ka and SP94Q (Gundersen et al. 1995; GRGS)\footnote{
Note that our analysis of the full SP94 data set accounts for all the
correlations.}, 
the ARGO Hercules
observations (de Bernardis et al. 1994; R99), the MAX 4 $\iota$ Draconis
(ID) and $\sigma$ Herculis (SH) and MAX 5 HR5127 (HR), $\mu$ Pegasi (MP),
and $\phi$ Herculis (PH) observations (Tanaka et al. 1996; Lim et al. 
1996; Ganga et al. 1998), the White Dish observations (Tucker et al. 1993;
Ratra et al. 1998), and the SuZIE observations (Church et al. 1997; Ganga
et al. 1997b). Detailed information about these data sets may be found in the 
papers cited above; in what follows we discuss only directly relevant
issues.

Amongst the experiments we consider, the SuZIE observations probe the 
smallest angular scales. In some models SuZIE is still quite sensitive
to multipoles $l \sim 2000$ (Ganga et al. 1997b). On these angular scales
a number of effects not accounted for in our CMB anisotropy power spectra
computations can modify the primordial CMB anisotropy power spectra. We 
therefore consider combined data sets both including and excluding the 
SuZIE observations, so as to bias our conclusions as little as possible. 

Multifrequency data sets, such as SP94 and MAX 4 and 5, allow for an
``internal" estimate of the amount of non-CMB anisotropy foreground 
contamination. In this sense they are ``better" than single frequency 
data sets.

Besides the well-known spectral analysis method for checking for
foreground contamination, some multifrequency data sets allow one to use
sample variance and noise considerations to check for consistency of the 
data with CMB anisotropy (GRGS, pp. 19--21). The method is as follows;
see GRGS for a detailed discussion and GRGS and Ganga et al. (1998) for
applications of the method. If systematic uncertainties (such as those in
the beamwidth, calibration, or pointing) are small, the deduced $Q_{\rm rms-PS}$
error bars only account for noise and sample variance. Assuming that the 
data is pure CMB anisotropy, when 
individual channel observations are combined, the noise contribution to 
the error bars will integrate down in known fashion while the sample 
variance contribution will not (since the sky coverage does not change).
Since the behavior of the noise part of the error bars is known, comparing
the error bars derived from each of the individual channel observations to
those derived from the combined data allows for an estimate of the sample
variance directly from the data. If this estimate of the sample variance 
is consistent with that estimated independently from the observing 
strategy and parameters of the instrument, it is reasonable to conclude 
that the data is not inconsistent with CMB anisotropy. 

Spectral analyses and simple analytic sample variance estimates indicate 
that the SP94Ka and MAX 5 HR data are more consistent with what is expected for 
CMB anisotropy than are the SP94Q and MAX 5 PH data (GRGS; Ganga et al. 1998, 
also see Tanaka et al. 1996)\footnote{
Given the error bars, the spectral analysis can not be used to argue that the
SP94Q data is not purely CMB anisotropy (GRGS).}. 
While the MAX 5 PH
observations were done at lower balloon altitude and so through more of
the atmosphere (Tanaka et al. 1996), and while there is indication of a possible
non-CMB foreground in other Q band observations made at Saskatoon 
(de Oliveira-Costa et al. 1997), it is useful to improve on the simple 
analytic sample variance estimates
of GRGS and Ganga et al. (1998) to see if their conclusions are justified.

To improve upon these estimates we have generated 1000 Monte Carlo maps 
of the flat bandpower CMB anisotropy in a flat two-dimensional space\footnote{
The SP94 and MAX 4 and 5 experiments probe a small enough region of the 
sky for the flat space approximation to hold.}. 
We then mimic the SP94 and MAX 4 and 5 observational techniques and 
``observe" these noise-less simulations. 

For the SP94 Ka and Q observations we find sample variances of 23\% and 22\% of 
the bandtemperature, estimated at the central beamwidths, as in the GRGS sample
variance estimate. These are larger than the 21\% and 18\% estimated by GRGS
on the basis of the simple analytic formula. However, the conclusions of 
GRGS are unaffected. That is, given these new numerical estimates of sample
variance, the SP94Ka data is still more consistent with what is
expected for CMB anisotropy than is the SP94Q data.

For MAX 5 we find a sample variance of 23\%, estimated at 
beamwidth $\sigma_{\rm fwhm} = 0.5^\circ$, as in the Ganga et al. (1998)
sample variance estimate. This is larger than the Ganga et al. (1998)
analytic estimate of 19\%. However, this is not large enough to affect the 
Ganga et al. (1998) conclusion that the MAX 5 HR error bars are more 
consistent with what is expected for CMB anisotropy than are the MAX 5 PH 
ones. We also note that MAX 5 MP data has structure that correlates with 
$IRAS$ 100 $\mu$m dust emission (Lim et al. 1996).
Since the MAX 4 SH error bars do not significantly shrink when the 
individual channel observations are combined, while the MAX 4 ID error 
bars do, Ganga et al. (1998) concluded that the MAX 4 SH data was
not that consistent with what is expected for CMB anisotropy, while MAX 4 ID
could be. 

On the basis of these, admittedly weak, arguments, we therefore also analyze 
the combination of SP94Ka, MAX 4 ID, and MAX 5 HR data. These are 
multifrequency observations with error bars consistent with what is 
expected for CMB anisotropy on the basis of sample variance and noise 
considerations. More precisely, these three data sets are the only ones 
(amongst those considered in this paper) for which we have no basis to suspect 
foreground contamination.\footnote{While there are notable exceptions, we
emphasize that other data at similar angular scales tend to have a larger 
amplitude.}
In this sense
this is a conservative combination of data. We emphasize, however, that some 
or all of the other data sets could also be purely CMB anisotropy, since
our arguments do not establish the contrary at a convincing statistical level.

We emphasize that a proper treatment would require a complete analysis of each 
individual data set, accounting for a CMB anisotropy signal as well as 
various non-CMB foreground signals. Since there almost certainly are 
uncharacterized non-CMB foregrounds, this is likely to be a
difficult undertaking.

In summary, then, we consider three different combinations of data sets: 
i) all data (SP94, ARGO, MAX 4 and 5, White Dish, and SuZIE); ii) all
data excluding SuZIE; and iii) SP94Ka, MAX 4 ID, and MAX 5 HR data.

\section{Summary of Computation}

GRGS describe the computation of the likelihood function for a 
given CMB anisotropy data set. Offsets and gradients removed from 
the data are accounted for in the analysis. Beamwidth and calibration
uncertainties are also accounted for as described in GRGS.

The likelihood functions for the individual data sets considered in this 
paper were recomputed in R99, for the much larger set of cosmogonical
model spectra considered here. The initial models-based analyses, except
for ARGO, used only 25 model spectra (Ratra et al. 1997), in contrast to
the 798 considered here. Since the data sets considered are well separated
in space, and in some cases also in angular resolution, the likelihood
functions of the individual data sets are simply multiplied together to 
construct the likelihood function of the combined data. These open and flat-$\Lambda$ model 
likelihoods are a function of four parameters: $Q_{\rm rms-PS}$, $\Omega_0$, 
$\Omega_B h^2$, and $t_0$. Marginalized likelihood functions are derived by 
integrating over one or more of these parameters. We assume a uniform prior 
in the parameters integrated over, set to zero outside the parameter range 
considered.

To determine central values and limits from the likelihood functions we 
assume a uniform prior in the relevant parameter. Then the corresponding 
posterior probability density distribution function vanishes outside the
chosen parameter range and is equal to the likelihood function inside
this range. The deduced central value of the parameter is taken to be the 
value at which the posterior probability density peaks, and we quote highest 
posterior density (HPD) limits. See GRGS and R99 for details. The 
quoted limits depend on the prior range considered for the parameter.
This is a significant effect if the likelihood function is not sharply peaked 
within the parameter range considered, as is the case for a number of the 
likelihood functions derived in this paper. See R99 for a more detailed 
discussion of this issue.

\section{Results and Discussion}

For the flat bandpower spectrum, the posterior distribution for the 
combined data peaks at $Q_{\rm rms-PS} = 25\ \mu$K, with a 1 
$\sigma$ range of $23\ \mu{\rm K} < Q_{\rm rms-PS} < 27\ \mu{\rm K}$, 
resulting in an averaged fractional 1 $\sigma$ uncertainty of 8.9\%, and 
with likelihood ratio\footnote{
This is the ratio of the value of the posterior distribution at the 
peak to that at $Q_{\rm rms-PS}$ = 0 $\mu$K.}
 = $ 4 \times 10^{99}$. For this spectrum, the posterior 
distribution for the combined data excluding SuZIE peaks at 
$Q_{\rm rms-PS} = 26\ \mu$K, with 1 $\sigma$ range of $24\ \mu{\rm K} < 
Q_{\rm rms-PS} < 29\ \mu{\rm K}$, resulting in an averaged fractional  
uncertainty of 9.5\%, and with likelihood ratio = $ 7 \times 10^{99}$.
For the SP94Ka, MAX 4 ID, and MAX 5 HR data combination, the posterior
distribution peaks at $Q_{\rm rms-PS} = 22\ \mu$K, with 1 $\sigma$ range of 
$18\ \mu{\rm K} < Q_{\rm rms-PS} < 26\ \mu{\rm K}$, resulting in an averaged 
fractional uncertainty of 19\%, and with likelihood ratio = $ 1 \times 10^{31}$.
These numerical values account for beamwidth and calibration uncertainties,
and the removal of offsets and gradients when appropriate. Clearly, these
combined data sets result in very significant detections of CMB anisotropy, 
even after known systematic uncertainties are accounted for. We emphasize that 
the most conservative combination (SP94Ka, MAX 4 ID, and MAX 5 HR) has a lower 
amplitude than the other two data combinations, but it is not significantly 
lower. Also, while this most conservative combination of data has a larger
error bar, 19\%, qualitatively consistent with a smaller amount of data, 
adding additional data to that considered here should result in a considerably 
smaller error bar. For comparison, the corresponding DMR error bar is 
$\sim 10-12\%$ (depending on model, G\'orski et al. 1998).

As for ARGO (R99), for both the open and flat-$\Lambda$ models, the 
four-dimensional posterior probability density distribution function 
$L(Q_{\rm rms-PS}, \Omega_0, \Omega_B h^2, t_0)$ is nicely peaked in the 
$Q_{\rm rms-PS}$ direction but fairly flat in the other three directions
(especially in the $\Omega_B h^2$ and $t_0$ directions).  Again as for
ARGO (R99), marginalizing over $Q_{\rm rms-PS}$ results in a three-dimensional 
posterior distribution $L(\Omega_0,  \Omega_B h^2, t_0)$ which is steeper.
Since the limits determined from the four- and three-dimensional posterior
distributions are not highly statistically significant, we do not show 
detailed contour plots of these functions here. However, the vertical heavy  
solid lines in Figure 8 show the formal 1 $\sigma$ and 2 $\sigma$ confidence 
limits derived
by projecting the appropriate four-dimensional posterior distribution
on to the $Q_{\rm rms-PS}$ axis. These limits are reasonably close 
together in the $Q_{\rm rms-PS}$ direction, illustrating the steepness of 
the four-dimensional posterior distribution in this direction.

Marginalizing over $Q_{\rm rms-PS}$ and one other parameter results in 
two-dimensional posterior probability functions which are more peaked. Some 
examples are shown in Figures 2--4. Again as for ARGO (R99), in some cases these
peaks are at an edge of the parameter range considered. Figures 2--4 
illustrate a number of interesting points, some of which we return to 
below when we discuss the posterior distribution obtained by marginalizing
over one more parameter, i.e., the posterior distributions for each of the 
four parameters.

Figure 4 shows results in the $(Q_{\rm rms-PS}, \ \Omega_0)$ plane for the 
open model and the individual MAX 4 ID, MAX 5 HR, and SP94Ka data sets, as 
well as the combined SP94Ka, MAX 4 ID, and MAX 5 HR data set. Panels $a)-c)$
show that the contours of the posterior distribution in the $\Omega_0$  
direction for these three individual data sets look like rescaled (in 
$Q_{\rm rms-PS}$) versions of 
each other. This is partly because the three experiments probe similar 
angular scale parts of the model CMB power spectra, the multipole 
$l \sim 50-100$ rise towards the first peak, which are only weakly dependent 
on $\Omega_0$ (see Figure 1), and partly because the data error bars are
large. As a result, when one does a joint analysis of the three data sets the 
contours in the $Q_{\rm rms-PS}$ direction are closer together while those in 
the $\Omega_0$ direction do not shift as significantly. Thus a combined 
analysis of the data sets results in tighter limits on $Q_{\rm rms-PS}$ but
not significantly stronger constraints on $\Omega_0$.
 
This effect has previously been noticed by GRGS, Bond \& Jaffe (1997), Ganga 
et al. (1998),
and R99. GRGS found that when the SP94 Ka and Q data were jointly analyzed the 
limits on $Q_{\rm rms-PS}$ were somewhat tighter than those derived from the 
Ka or Q data alone while the combined data did not better differentiate 
between different $\Omega_0$ values than did the individual data sets. 
Ganga et al. (1998) also drew attention to this effect in their
joint analysis of the MAX 4 and 5 data. Figure 5 of R99 shows posterior 
density distribution contours in the $(Q_{\rm rms-PS}, \ \Omega_0)$ plane
for the DMR, SP94, ARGO, MAX 4 and 5, White Dish, and SuZIE data. 

Bond \& Jaffe (1997) use the amplitude of mass fluctuations at 8$h^{-1}$ Mpc,
$\delta M/M (8 h^{-1}\ {\rm Mpc})$, instead of $Q_{\rm rms-PS}$, and show 
posterior density distribution contours in the $(\delta M/M (8 h^{-1}\ 
{\rm Mpc}), \ n)$ and $(\delta M/M (8 h^{-1}\ {\rm Mpc}), \ h)$ planes
instead of the plots we show. (Here $n$ is the primordial energy-density 
perturbation power spectral index.) They consider the combination of the DMR,
SP94, and Saskatoon data, and point out that since the DMR data does constrain
$n$ fairly well (G\'orski et al. 1996) but not $h$ (e.g., G\'orski et al. 1998),
the combination of DMR, SP94, and Saskatoon data does constrain $n$ fairly
well, but not $h$. A similar constraint on $\Omega_0$ in the open models
at low $\Omega_0$ (as that on $n$) would be expected from the SuZIE data 
(and to a lesser extent from the DMR data) --- see Figure 5 panels $h)$ and 
$b)$ of R99, and Figure 3 panels $b)$ and $d)$ of this paper.

Figure 2 shows that the two-dimensional posterior distributions allow one to 
distinguish between different regions of parameter space at a fairly
high formal level of confidence. For instance, for the SP94Ka, MAX 4 ID, and 
MAX 5 HR data combination (panels $c)$ and $f)$ in the bottom row), the open 
model near $\Omega_0 \sim 0.7$, $\Omega_B h^2 \sim 0.03$, and $t_0 \sim 20$ 
Gyr, and the flat-$\Lambda$ model near $\Omega_0 \sim 0.45$, $\Omega_B h^2
\sim 0.03$, and $t_0 \sim 20$ Gyr, are both formally ruled out at $\sim 3$ 
$\sigma$ confidence. However, we emphasize, as discussed below, care must be 
exercised when interpreting the discriminative power of these formal limits, 
since they depend sensitively on the fact that the uniform prior has been set 
to zero outside the range of the parameter space we have considered.

Figure 3 shows the contours of the two-dimensional posterior distribution 
for $Q_{\rm rms-PS}$ and $\Omega_0$, derived by marginalizing the 
four-dimensional distribution over $\Omega_B h^2$ and $t_0$. These are shown
for the three combined small-scale data sets and the DMR data, for 
both the open and flat-$\Lambda$ models. At 2 $\sigma$ confidence, constraints 
on these 
parameters derived from the three combined small-scale data sets are mostly 
consistent with those derived from the DMR data. However, at a lower level 
of significance, the DMR data favor a higher normalization for the 
flat-$\Lambda$ model than do the combined small-scale data sets, see panels 
$a)$, $c)$, and $e)$ of Figure 3. It is interesting that the constraints 
on $Q_{\rm rms-PS}$ derived from the DMR data are not much tighter than those
derived from the combined small-scale data sets. A similar conclusion has 
previously been drawn from the MAX 4 and 5 data alone (Ganga et al. 1998) 
and the ARGO data alone (R99). As mentioned above, excluding SuZIE from the 
combined small-scale data set results in an increase of the allowed 
low-$\Omega_0$ region for the open model, see panels $b)$ and $d)$ of Figure 3.

Figure 5 shows the one-dimensional posterior distribution functions for
$\Omega_0$, derived by marginalizing the 
four-dimensional ones over the other three parameters. For the flat-$\Lambda$
model (left hand column of Figure 5), all three combinations of data sets
favor $\Omega_0 = 1$, although no value of $\Omega_0$ is ruled out since the
distributions are quite flat. Including or excluding the SuZIE data, panels $a)$ and $c)$, hardly changes the posterior distribution for the flat-$\Lambda$ 
model. This is mostly a consequence of the fact that the flat-$\Lambda$ models 
do not predict much CMB anisotropy power on the angular scales probed by SuZIE
(see Figure 1).
The results for the open model (right hand column of Figure 5) are dramatically
different. Using all the data, panel $b)$, $\Omega_0 = 1$ is favored, 
with, formally, $\Omega_0 > 0.55,\ > 0.28,\ {\rm and} > 0.1$ at 1 $\sigma$, 
2 $\sigma$, and 3 $\sigma$ confidence, respectively. Excluding SuZIE, panel 
$d)$, significantly shifts the most likely value to $\Omega_0 = 0.28$. 
Considering only the SP94Ka, MAX 4 ID, 
and MAX 5 HR combination, panel $f)$, the most favored open model is
at $\Omega_0 = 0.1$ with  $0.47 < \Omega_0 < 0.87$ formally excluded at
1 $\sigma$ and $0.63 < \Omega_0 < 0.71$ formally excluded at 2 $\sigma$.

As mentioned above, and discussed in the analysis of the ARGO data (R99), 
care is needed when interpreting the discriminative power of these formal 
limits. If the posterior density function was a gaussian which peaked 
well inside the parameter range considered, the 1, 2, and 3 $\sigma$ HPD 
limits would correspond to a value of the posterior distribution relative to 
that at the peak of 0.61, 0.14, and 0.011 respectively. Using this criterion,
the posterior distribution of Figure 5$b)$, for the open model and all the 
small-scale data, sets a 2 $\sigma$ confidence limit of $\Omega_0 \gap 0.2$ 
and a 3 $\sigma$ confidence limit of $\Omega_0 \gap 0.1$. This gaussian
posterior ratio criterion also results in a 1 $\sigma$ confidence limit of
$\Omega_0 \lap 0.3$ for the open model and the SP94Ka, MAX 4 ID, and MAX 5 HR 
combination, panel $f)$. This criterion does not result in 1 $\sigma$ limits 
for any of the other panels.

To summarize, the most conservative constraints follow from the SP94Ka,
MAX 4 ID, and MAX 5 HR data combination, Figure 5 panels $e)$ and $f)$
--- the most favored value is $\Omega_0 = 0.1\ (1)$ for the open 
(flat-$\Lambda$) model, with $\Omega_0 \gap 0.3$ excluded at 1 $\sigma$
for the open model and no constraint on the flat-$\Lambda$ model if the
gaussian posterior distribution ratio criterion is used. 
That is, even using this most conservative prescription a 1 $\sigma$ limit
can be set on $\Omega_0$ in the open model, and the most favored open model 
is somewhat more favored than the most favored flat-$\Lambda$ one. 

While the inclusion or exclusion of the SuZIE data do not significantly
affect the flat-$\Lambda$ models constraints, panels $a)$ and $c)$ of Figure
5 or panels $a)$ and $c)$ of Figure 3, it does significantly affect the open 
model constraints, panels $b)$ and $d)$ of Figure 5 or panels $b)$ and $d)$ of 
Figure 3, shifting the most likely value of $\Omega_0$ from 1 
to 0.3. Since the SuZIE data only results in an upper limit
(Church et al. 1997; Ganga et al. 1997b), it is not straightforward to 
include it in a conventional goodness-of-fit $(\chi^2)$ analysis. This is 
a drawback of the conventional $\chi^2$ technique.

It is of interest to compare the constraints on $\Omega_0$
in the open model from the combined SP94Ka, MAX 4 ID, and MAX 5 HR 
analysis\footnote{
We again emphasize that these results are tentative since we consider only 
three data sets. Also, as noted above, other data at similar angular scales 
tend to have a larger amplitude.}
with what has been derived from $\chi^2$ analyses which include much more data.
As mentioned above, we find, from Figure 5$f)$, that $\Omega_0 = 0.1$ is the 
most favored value, with the conservative gaussian posterior distribution ratio
upper limit of $\Omega_0 \lap 0.3$ at 1 $\sigma$ and no constraint at 
2 $\sigma$. This is not inconsistent with the Ganga et al. (1996) $\chi^2$ 
analysis conclusion that $\Omega_0 \sim 0.4-0.5$ is favored. It is also not 
inconsistent with the Baker et al. (1998) values $\Omega_0 = 0.7 
{+0.6 \atop -0.4}$ at 2 $\sigma$, where we have simply doubled their 1 $\sigma$ 
error bars.  However, it seems to be difficult to reconcile this with the 
Lineweaver \& Barbosa (1998) conclusion that $\Omega_0 > 0.3$ at more
than $\sim 4$ $\sigma$ confidence (if $h = 0.7$). This difference could be 
due to a number of effects. It could be due to the fact that we have considered only a small number of data sets. Another possibility is the different numerical values ascribed to some of the data sets in the two analyses. For 
instance, recent reanalyses (GRGS; Ganga et al. 1997b, 1998; Ratra et al. 1998; 
R99) of some of the data, which more carefully accounts for systematic effects, 
has altered a number of data points, some quite significantly. Yet another 
possibility is that some of the data sets considered in the $\chi^2$ analysis
but not in this paper are inconsistent with the data sets considered here, for
the restricted four-dimensional parameter space we have studied. This, however, 
does not seem very likely since $\chi^2$ analyses have resulted in very low 
reduced $\chi^2$ values for some of the models considered here, indicating that 
in this restricted parameter space the data sets are qualitatively consistent 
with each other given the error bars (Ganga et al. 1996; Lineweaver \& 
Barbosa 1998; Tegmark 1998). Given that the field is still rapidly evolving, 
such differences are not unexpected. 
 
Figure 6 shows the one-dimensional posterior distribution functions for 
$\Omega_B h^2$, derived by marginalizing the four-dimensional ones over the 
other three parameters. For both the flat-$\Lambda$ and open models, and 
for all combined data sets considered, a low $\Omega_B h^2 \simeq 0.005$ is
favored. While formal 2 $\sigma$ confidence upper limits exist (see 
Figure 6), the constraints derived from the gaussian posterior distribution
ratio criterion discussed above are much weaker and do not rule out any value
of $\Omega_B h^2$. Inclusion or exclusion of the SuZIE data do not 
significantly affect the deductions in this case.
 
While not very significant, it is gratifying that the CMB anisotropy 
data favor lower $\Omega_B h^2$, consistent with a number of recent estimates 
using different techniques, which disfavor $\Omega_B h^2$ larger than 
$\sim 0.02$ (e.g., Fukugita, Hogan, \& Peebles 1998; Levshakov, Kegel, \& 
Takahara 1998, but also see Burles \& Tytler 1998). This low value for 
$\Omega_B h^2$ is also consistent with the indications from the Ganga et al. (1996) $\chi^2$ analysis. However, it differs from the Lineweaver \& Barbosa 
(1998) favored high value of $\Omega_B h^2 \simeq 0.026$.

Figure 7 shows the one-dimensional posterior distribution functions for 
$t_0$, derived by marginalizing the four-dimensional ones 
over the other three parameters. For both the flat-$\Lambda$ and open models, 
and for all combined data sets considered, a young universe, $t_0 \sim 
10-12$ Gyr is favored. Again, while formal 2 $\sigma$ confidence upper limits 
exist (see Figure 7), constraints derived from the gaussian posterior 
distribution ratio criterion are much weaker, although they are stronger than 
for the case of $\Omega_B h^2$. These 1 $\sigma$ constraints are:
for all the data and the open model, panel $b)$, $t_0 \lap 16$ Gyr;
for all the data excluding SuZIE and the open model, panel $d)$, 
$t_0 \lap 19.5$ Gyr; and for the SP94Ka, MAX 4 ID, and MAX 5 HR combination,
panels $e)$ and $f)$, $t_0 \lap 13.5$ Gyr ($t_0 \lap 15.5$ Gyr), for the 
flat-$\Lambda$ (open) models. Inclusion or exclusion of the SuZIE data
do not dramatically affect the deductions in this case.

Again, while not highly significant, it is gratifying that the CMB anisotropy 
data favor a young universe, consistent with other recent estimates
(e.g., Feast \& Catchpole 1997; Reid 1997; Gratton et al. 1997; Chaboyer et al. 
1998). Lower $t_0$ corresponds to larger $h$. Hence these results would seem to 
be consistent with the indications for large $h$ from the Ganga et al. (1996) 
$\chi^2$ analysis. They also seem to be consistent 
with the open model values of $h \sim 0.6$ favored by Lineweaver \& Barbosa 
(1998) and Baker et al. (1998). However, they would seem to be difficult
to reconcile with the flat-$\Lambda$ model value of $h \sim 0.35$ 
favored by Baker et al. (1998).

Figure 8 shows the one-dimensional posterior distribution functions for
$Q_{\rm rms-PS}$, derived by marginalizing the four-dimensional 
ones over the other three parameters. As discussed above, the three combined
data sets result in fairly tight constraints on $Q_{\rm rms-PS}$. At 
2 $\sigma$ confidence they are consistent with the DMR results for both the 
open and flat-$\Lambda$ models. However, at a lower level of confidence the DMR 
data favor a somewhat higher normalization for the flat-$\Lambda$ model 
than do the small-scale data sets. This is consistent with indications from
other data (e.g., Stompor et al. 1995; Liddle et al. 1996; Cole et al. 1997). 

The peak values of the one-dimensional posterior distributions shown in 
Figures 5--8 are listed in the figure captions for the case when the 
four-dimensional posterior distributions are normalized such that
$L(Q_{\rm rms-PS}\ =\ 0\ \mu{\rm K})\ =\ 1$. With this normalization, 
marginalizing over the remaining parameter the fully marginalized
posterior distributions are, for the open (flat-$\Lambda$) models:
$5\times 10^{100}(9\times 10^{100})$ for all the data;   
$4\times 10^{100}(4\times 10^{100})$ for all the data excluding SuZIE;
and, $6\times 10^{30}(6\times 10^{30})$ for the SP94Ka, MAX 4 ID, and 
MAX 5 HR combination.

\section{Conclusion}

We have derived constraints on cosmological model parameters in the open and 
flat-$\Lambda$ CDM models, from joint analyses of combinations of the SP94, 
ARGO, MAX 4+5, White Dish, and SuZIE CMB anisotropy data sets.

The favored value of $\Omega_0$ in 
the open model depends sensitively on which combination of data is used. 
To determine if this is significant will require a models-based analysis of a 
larger number of data sets. Constraints on $\Omega_0$ in the flat-$\Lambda$ 
model, and constraints on $\Omega_B h^2$ and $t_0$ in both models, are only 
weakly dependent on the data set combination considered.

In the most conservative case, i.e., for the SP94Ka, MAX 4 ID, and MAX 5 HR 
data set combination, an open (flat-$\Lambda$) model with $\Omega_0 \simeq 
0.1\ (1)$ is favored. We emphasize that, using the gaussian posterior ratio
criterion, flat-$\Lambda$ models with $\Omega_0$ in the range from 0.1 to
1 cannot be ruled out at even 1 $\sigma$ confidence, although open models
require $\Omega_0 \lap 0.3$ at 1 $\sigma$ confidence. 

For both the open and flat-$\Lambda$ models, and all data set combinations
considered, low $\Omega_B h^2 \sim 0.005$, or young, $t_0 \sim 10$ Gyr,
universes are weakly favored. This is in agreement with recent determinations
based on other techniques.

It is gratifying that the small subset of current CMB anisotropy data 
considered in this paper can place some constraints on cosmological parameters.
While these are mostly not very statistically significant, they are largely 
consistent with results determined using other, non-CMB anisotropy, 
measurements. To derive tighter and more robust constraints on cosmological 
parameters will require a models-based combined analysis of a much larger 
collection of CMB anisotropy data sets, including DMR.

\bigskip

We acknowledge useful discussions with C. Kochanek. We thank C. Lineweaver 
for a prompt, detailed and useful referee report, which helped improve the 
manuscript. This work was partially
carried out at the Infrared Processing and Analysis Center and the Jet
Propulsion Laboratory of the California Institute of Technology, under a
contract with the National Aeronautics and Space Administration. KG also
acknowledges support from NASA ADP grant NASA-1260. RS acknowledges support 
from a UK PPARC grant and from Polish Scientific Committee (KBN) grant 
2P03D00813.
BR and GR acknowledge support from NSF grant EPS-9550487 with matching support 
from the state of Kansas and from a K$^*$STAR First award. 



\clearpage
\centerline{\bf Figure Captions}

\medskip
\noindent
Fig.~1.--
{\protect CMB anisotropy multipole moments $l(l+1)C_l/(2\pi )\times
    10^{10}$ as a function of
    multipole $l$, for selected models normalized to the DMR data
    (G\'orski et al. 1998; Stompor 1997). Panels $a)-c)$ show selected
    flat-$\Lambda$ models. The heavy lines are the $\Omega_0 = 1$,
    $\Omega_B h^2 = 0.005$, and $t_0 = 10$ Gyr case, which is close to 
    where the SP94Ka, MAX 4 ID, and MAX 5 HR data set combination posterior 
    density distributions (marginalized over all but one parameter at a time)  
    are at a maximum. Panel $a)$ shows five $\Omega_B h^2$ = 0.005,
    $t_0$ = 10 Gyr models with $\Omega_0$ = 0.2, 0.4, 0.6, 0.8, and 1
    in descending order at the $l \sim 200$ peaks. Panel $b)$ shows seven 
    $\Omega_0$ = 1, $t_0$ = 10 Gyr models with $\Omega_B h^2$ = 0.029,
    0.025, 0.021, 0.017, 0.013, 0.009, and 0.005 in descending order at
    the $l \sim 200$ peaks. Panel $c)$ shows six $\Omega_0$ = 1, 
    $\Omega_B h^2$ = 0.005 models with $t_0$ = 20, 18, 16, 14, 12, and
    10 Gyr in descending order at the $l \sim 200$ peaks. Panels $d)-f)$ 
    show selected open models. The heavy lines are the $\Omega_0 = 0.1$,
    $\Omega_B h^2 = 0.005$, and $t_0 = 10$ Gyr case, which is close to 
    where the SP94Ka, MAX 4 ID, and MAX 5 HR data set combination posterior 
    density distributions (marginalized over all but one parameter at
    a time) are at a maximum. Panel $d)$ shows five $\Omega_B h^2$ = 0.005,
    $t_0$ = 10 Gyr models with $\Omega_0$ = 0.9, 0.7, 0.5, 0.3, and 0.1
    from left to right at the peaks. Panel $e)$ shows seven 
    $\Omega_0$ = 0.1, $t_0$ = 10 Gyr models with $\Omega_B h^2$ = 0.029,
    0.025, 0.021, 0.017, 0.013, 0.009, and 0.005 in descending order at
    $l \sim 600$. Panel $f)$ shows six $\Omega_0$ = 0.1, $\Omega_B h^2$ 
    = 0.005 models with $t_0$ = 20, 18, 16, 14, 12, and 10 Gyr in descending 
    order at $l \sim 600$.}
 
\medskip
\noindent
Fig.~2.--
{\protect Confidence contours and maxima of the two-dimensional posterior 
   probability density distribution functions, as a function of the two 
   parameters on the axes of each panel (derived by marginalizing the 
   four-dimensional posterior distributions over the other two parameters). 
   Dashed lines (crosses) show the contours (maxima) of the open case and 
   solid lines (solid circles) show those of the flat-$\Lambda$ model. 
   Contours of 0.25, 0.5, 1, 2, and 3 $\sigma$ confidence are shown
   (some contours are not labelled). Panels $a)-c)$ in the left 
   column show the $(\Omega_B h^2,\ \Omega_0)$ plane, while panels 
   $d)-f)$ in the right column show the $(t_0, \ \Omega_0)$ plane. Panels
   $a)$ \&\ $d)$ in the top row are from an analysis of all the small 
   scale data. Panels $b)$ \&\ $e)$ in the middle row are from an analysis of   
   all but the SuZIE small scale data. Panels $c)$ \&\ $f)$ in the bottom row 
   are from an analysis of the SP94Ka, MAX 4 ID, and MAX 5 HR data sets.}
  
\medskip
\noindent
Fig.~3.-- 
{\protect Confidence contours and maxima of the two-dimensional 
   $(Q_{\rm rms-PS}, \Omega_0)$ posterior probability density distribution 
   functions. Panels $a)$, $c)$, \&\ $e)$ in the left column show the 
   flat-$\Lambda$ model and panels $b)$, $d)$, \&\ $f)$ in the right column 
   show the open model. Note the different scale on the vertical 
   $(Q_{\rm rms-PS})$ axes of pairs of panels in each row. Shaded regions 
   and those enclosed by thin solid lines show  
   the results derived from the various combination data sets considered.
   Densest shading shows the 1 $\sigma$ confidence region, less-dense shading 
   shows the 2 $\sigma$ region, and thin solid lines enclose the 3 $\sigma$
   region. Solid circles show the maxima of the two-dimensional posterior 
   distributions. Heavy lines show the two-dimensional posterior probability 
   density distribution function 1 and 2 $\sigma$ confidence limits for the 
   DMR data (G\'orski et al. 1998; Stompor 1997). The DMR results are a 
   composite of 
   those from analyses of the two extreme data sets: i) galactic frame with 
   quadrupole included and correcting for faint high-latitude galactic 
   emission; and ii) ecliptic frame with quadrupole excluded and no other 
   galactic emission correction (G\'orski et al. 1998). Panels
   $a)$ \&\ $b)$ in the top row are from an analysis of all the small 
   scale data. Panels $c)$ \&\ $d)$ in the middle row are from an analysis of   
   all but the SuZIE small scale data. Panels $e)$ \&\ $f)$ in the bottom row 
   are from an analysis of the SP94Ka, MAX 4 ID, and MAX 5 HR data sets.}
  
\medskip
\noindent
Fig.~4.--
{\protect Confidence contours and maxima of the two-dimensional 
   $(Q_{\rm rms-PS}, \Omega_0)$ posterior probability density distribution 
   functions for the open model. Conventions are as described in the 
   caption of Figure 3. Panel $a)$ is from an analysis of the MAX 4 ID data, 
   panel $b)$ from the MAX 5 HR data, and panel $c)$ from the SP94Ka data.
   Panel $d)$ is from a combined  analysis of the SP94Ka, MAX 4 ID, and 
   MAX 5 HR data sets. Note the different scale on the vertical 
   $(Q_{\rm rms-PS})$ axis in panel $a)$.}

\medskip
\noindent
Fig.~5.--
{\protect One-dimensional posterior probability density distribution 
   functions for $\Omega_0$, derived by marginalizing the four-dimensional 
   ones over the other three parameters, in the open and flat-$\Lambda$ models.
   These have been renormalized to unity at the peaks. Dotted lines show the 
   formal 1 and 2 $\sigma$ confidence limits derived from these 
   one-dimensional posterior distributions.
   Panels $a)$, $c)$, \&\ $e)$ in the left column show the flat-$\Lambda$ model 
   and panels $b)$, $d)$, \&\ $f)$ in the right column show the open model. 
   Panels $a)$ \&\ $b)$ in the top row are from an analysis of all the small 
   scale data. Panels $c)$ \&\ $d)$ in the middle row are from an analysis of   
   all but the SuZIE small scale data. Panels $e)$ \&\ $f)$ in the bottom row 
   are from an analysis of the SP94Ka, MAX 4 ID, and MAX 5 HR data sets.
   When the four-dimensional posterior distributions are normalized such
   that $L(Q_{\rm rms-PS}\ =\ 0\ \mu{\rm K})\ =\ 1$, the peak values of
   the one-dimensional distributions shown in panels $a)-f)$ are 
   $1\times 10^{101}$, $1\times 10^{101}$, $4\times 10^{100}$, 
   $6\times 10^{100}$, $6\times 10^{30}$, and $1\times 10^{31}$,
   respectively.}

\medskip
\noindent
Fig.~6.--
{\protect One-dimensional posterior probability density distribution 
   functions for $\Omega_B h^2$, derived by marginalizing the four-dimensional 
   ones over the other three parameters, in the open and flat-$\Lambda$ models.
   Conventions are as described in the caption of Figure 5. When the  
   four-dimensional posterior distributions are normalized such
   that $L(Q_{\rm rms-PS}\ =\ 0\ \mu{\rm K})\ =\ 1$, the peak values of
   the one-dimensional distributions shown in panels $a)-f)$ are 
   $4\times 10^{102}$, $3\times 10^{102}$, $2\times 10^{102}$, 
   $2\times 10^{102}$, $3\times 10^{32}$, and $3\times 10^{32}$,
   respectively.}   
  
\medskip
\noindent
Fig.~7.--
{\protect One-dimensional posterior probability density distribution 
   functions for $t_0$, derived by marginalizing the four-dimensional 
   ones over the other three parameters, in the open and flat-$\Lambda$ models.
   Conventions are as described in the caption of Figure 5. When the  
   four-dimensional posterior distributions are normalized such
   that $L(Q_{\rm rms-PS}\ =\ 0\ \mu{\rm K})\ =\ 1$, the peak values of
   the one-dimensional distributions shown in panels $a)-f)$ are 
   $1\times 10^{100}$, $8\times 10^{99}$, $4\times 10^{99}$, 
   $5\times 10^{99}$, $1\times 10^{30}$, and $8\times 10^{29}$,
   respectively.}   
  
\medskip
\noindent
Fig.~8.--
{\protect One-dimensional posterior probability density distribution functions 
   for $Q_{\rm rms-PS}$, derived by marginalizing the four-dimensional 
   ones over the other three parameters, in the open and flat-$\Lambda$ models.
   Conventions are as described in the caption of Figure 5. Also shown here, 
   as dotted 
   lines, are the formal 3 $\sigma$ confidence limits derived from these 
   one-dimensional posterior distributions. Solid vertical lines show the 
   $\pm 1$ and $\pm 2$ $\sigma$ confidence limits derived by projecting the 
   corresponding small-scale data four-dimensional posterior distributions. 
   Also shown are the 2 $\sigma$ DMR (marginalized and projected) confidence 
   limits; these are a composite of those from the two extreme DMR data sets 
   (see caption of Figure 3). When the  
   four-dimensional posterior distributions are normalized such
   that $L(Q_{\rm rms-PS}\ =\ 0\ \mu{\rm K})\ =\ 1$, the peak values of
   the one-dimensional distributions shown in panels $a)-f)$ are 
   $2\times 10^{100}$, $9\times 10^{99}$, $8\times 10^{99}$, 
   $4\times 10^{99}$, $8\times 10^{29}$, and $4\times 10^{29}$,
   respectively.}

\clearpage
\pagestyle{empty}

\begin{center}
  \leavevmode
  \epsfysize=8.0truein
  \epsfbox{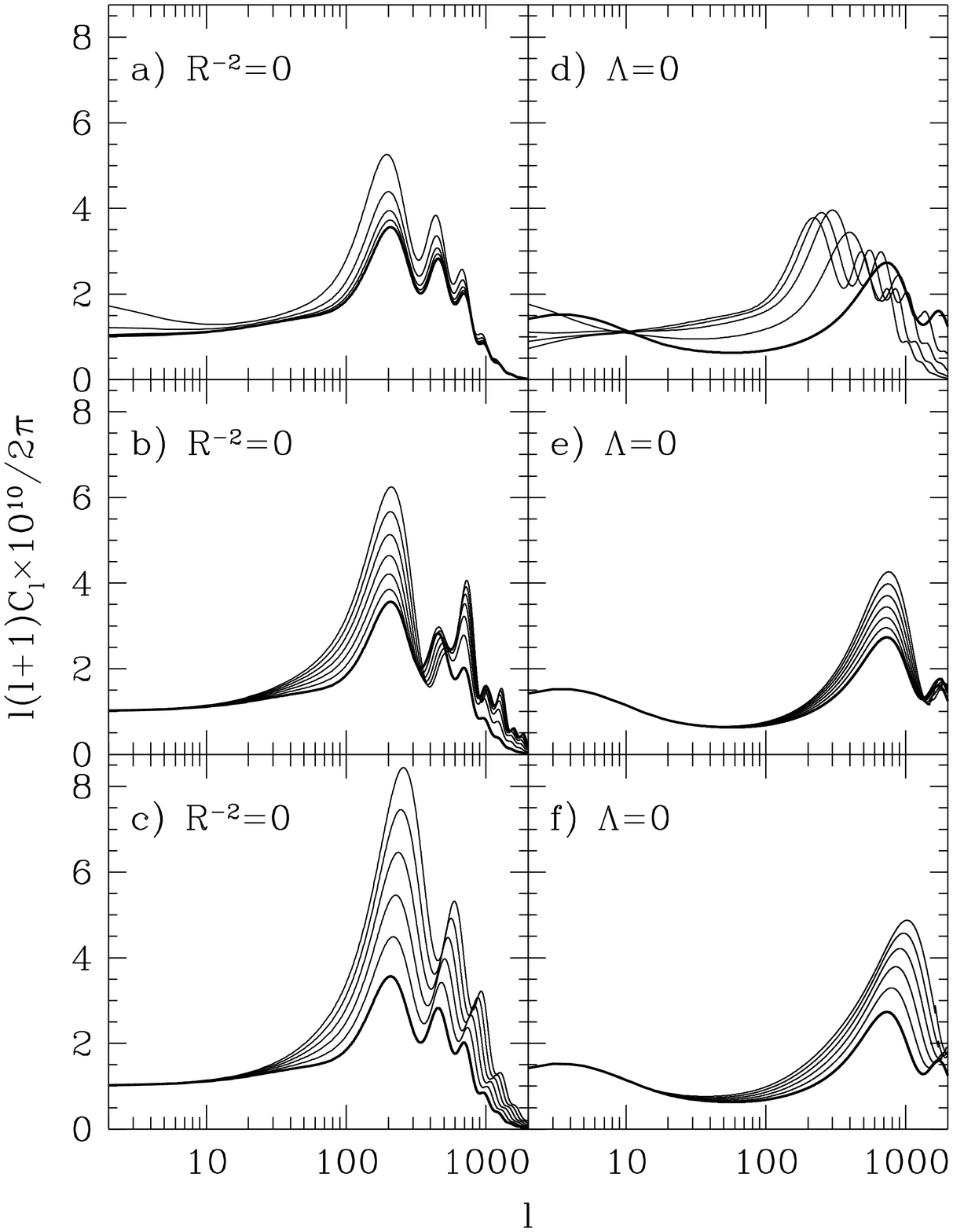}
\end{center}  
\vfill
Figure 1

\clearpage
\begin{center}
  \leavevmode
  \epsfysize=8.0truein
  \epsfbox{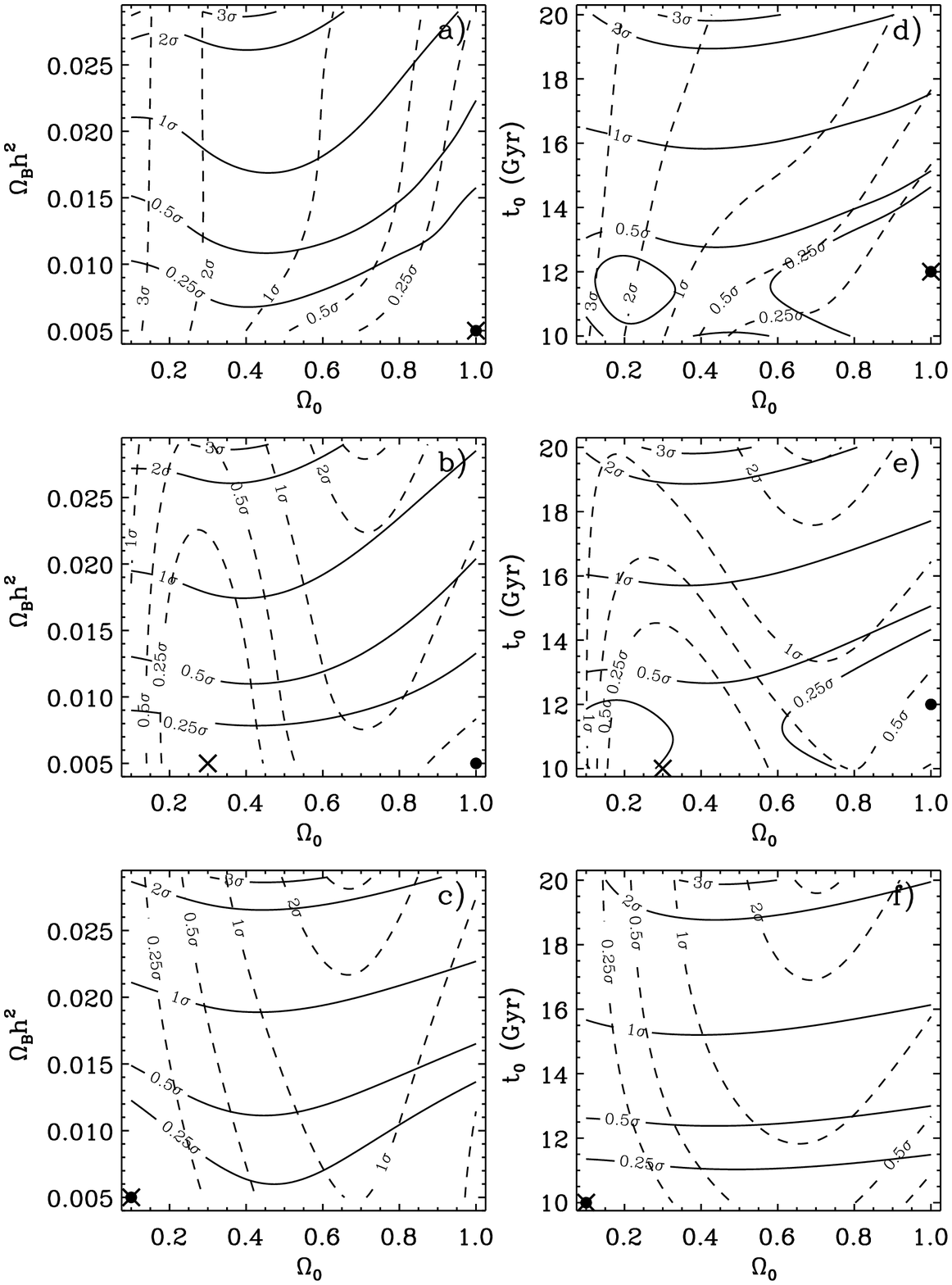}
\end{center}  
\vfill
Figure 2

\clearpage
\begin{center}
  \leavevmode
  \epsfysize=8.0truein
  \epsfbox{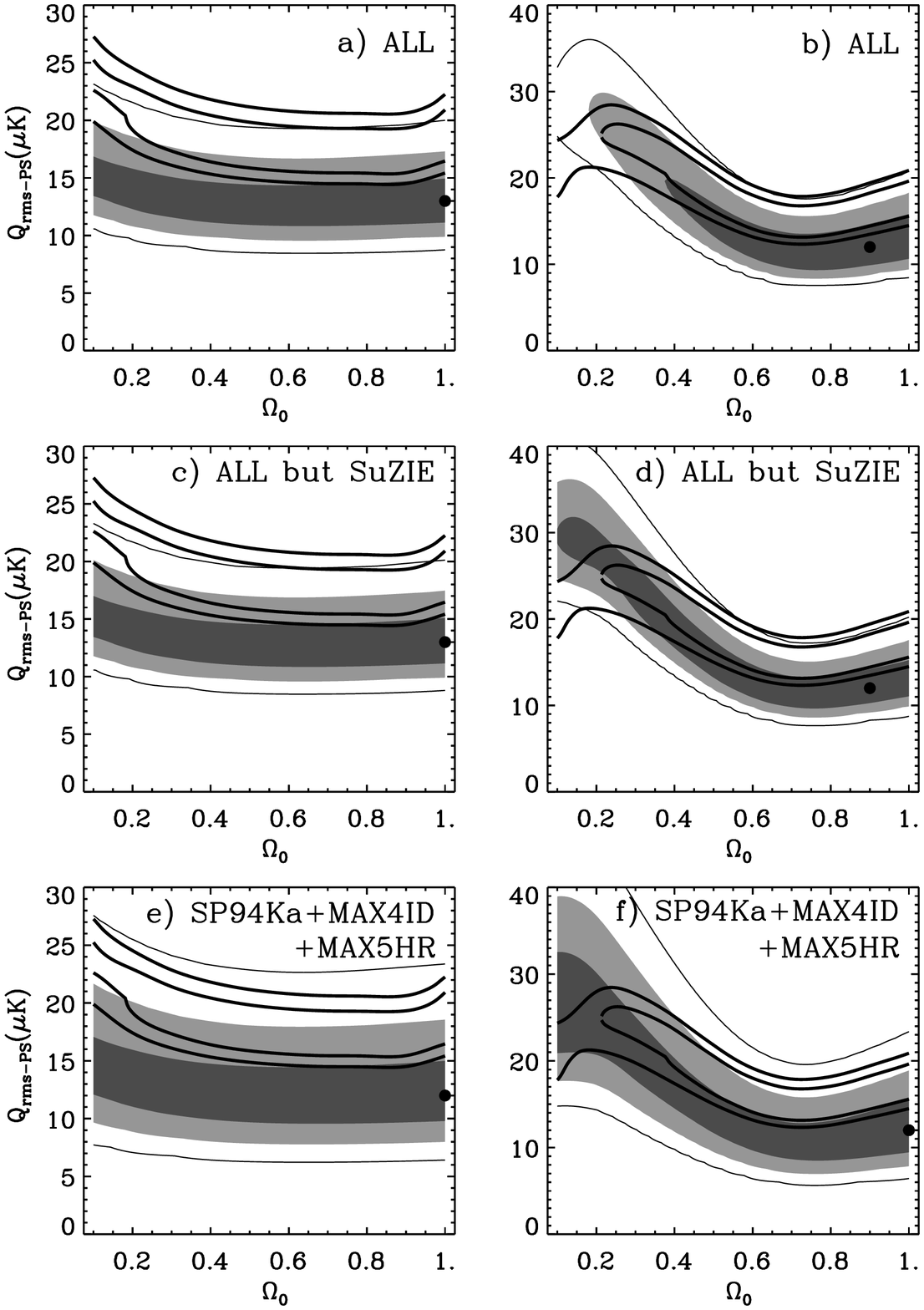}
\end{center}  
\vfill
Figure 3

\clearpage
\begin{center}
  \leavevmode
  \epsfysize=8.0truein
  \epsfbox{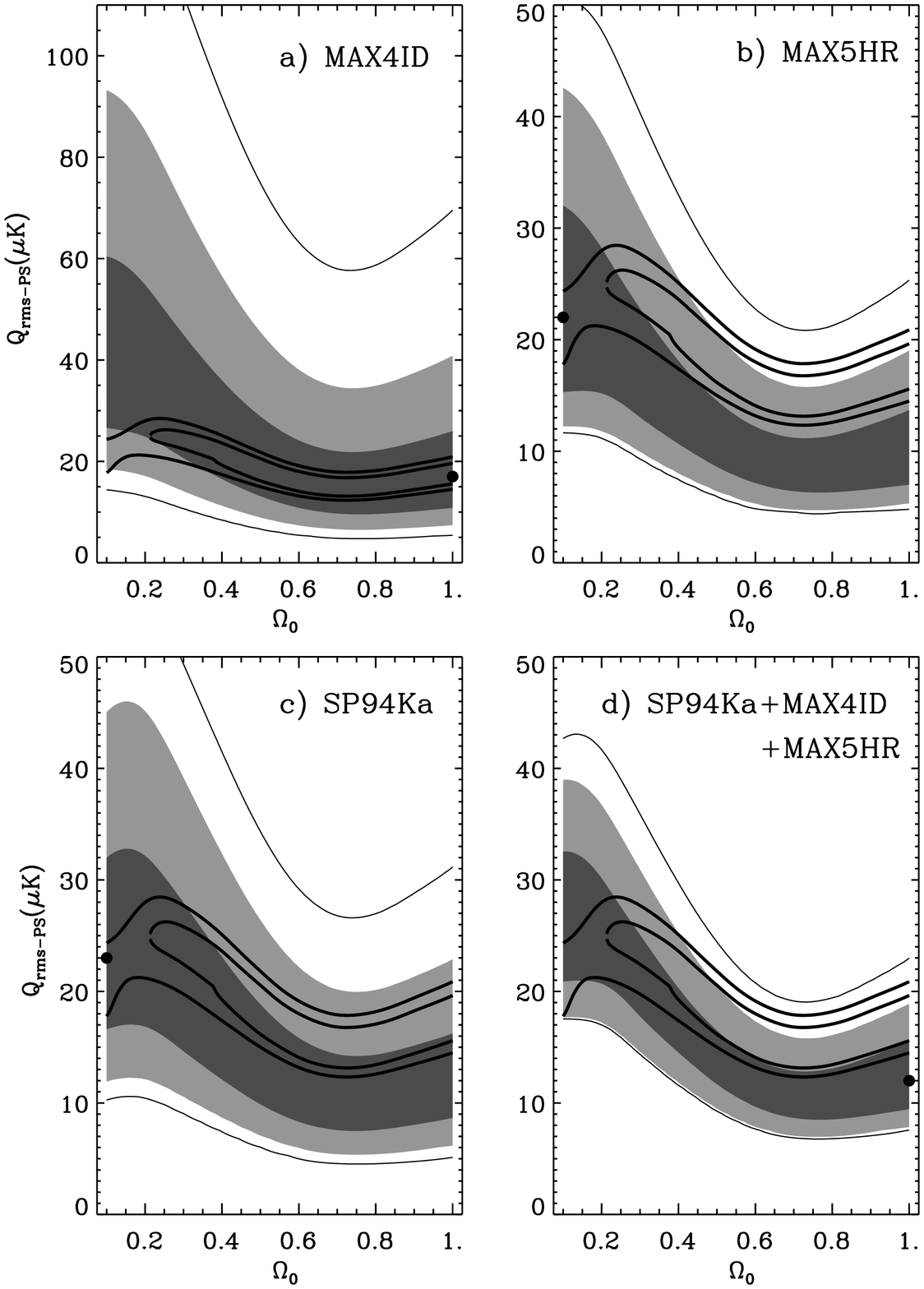}
\end{center}  
\vfill
Figure 4

\clearpage
\begin{center}
  \leavevmode
  \epsfysize=8.0truein
  \epsfbox{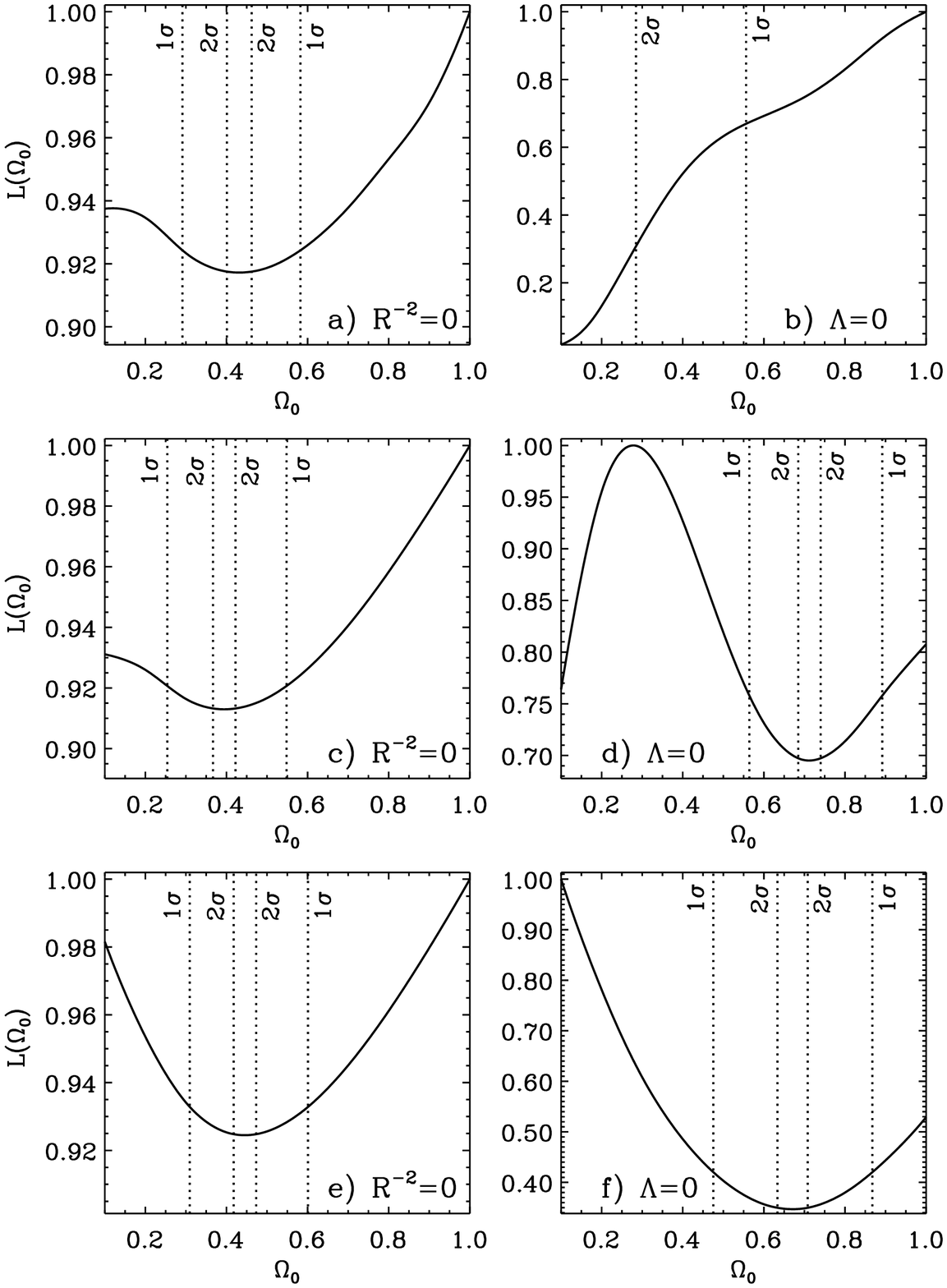}
\end{center}  
\vfill
Figure 5

\clearpage
\begin{center}
  \leavevmode
  \epsfysize=8.0truein
  \epsfbox{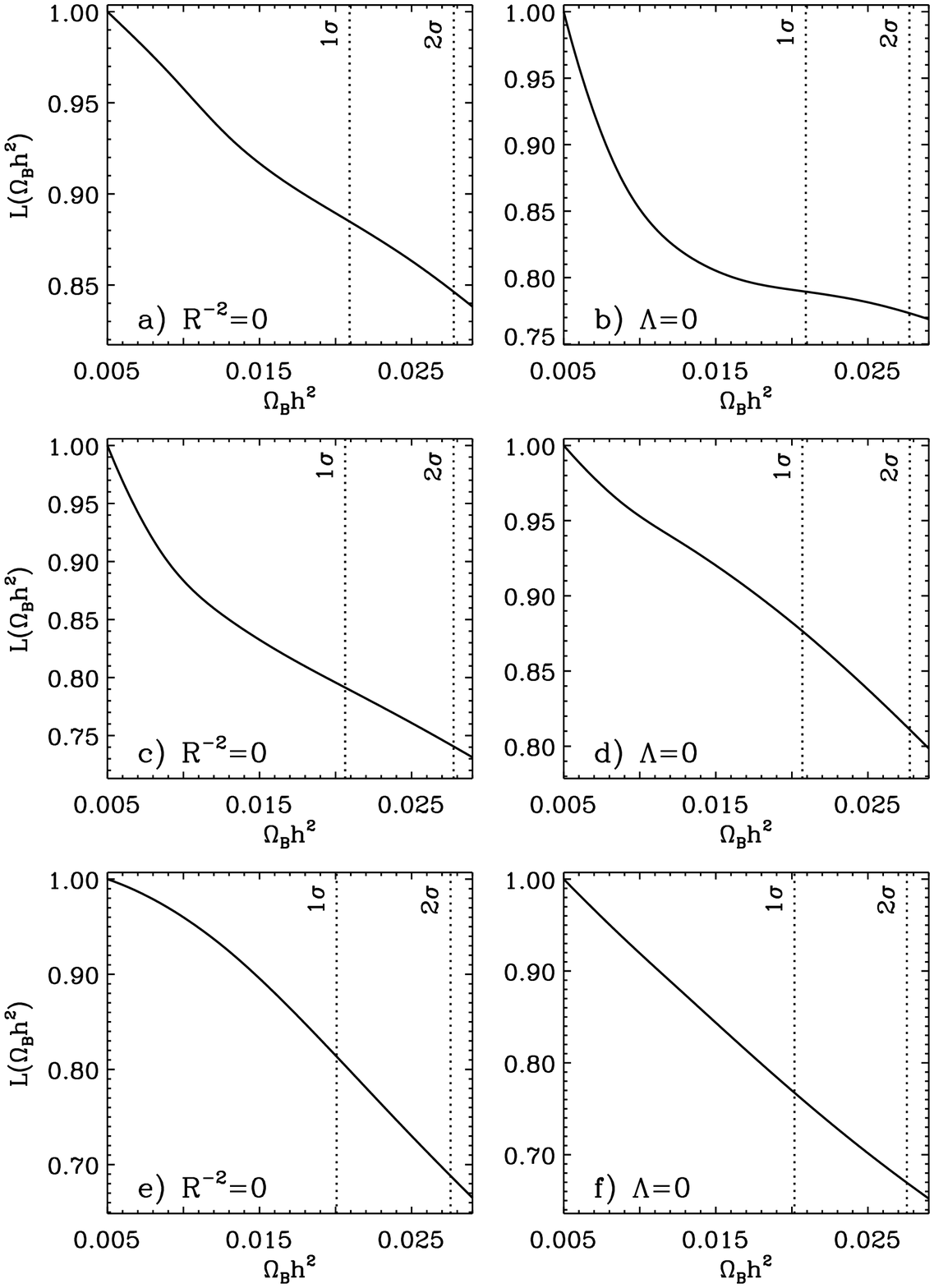}
\end{center}  
\vfill
Figure 6

\clearpage
\begin{center}
  \leavevmode
  \epsfysize=8.0truein
  \epsfbox{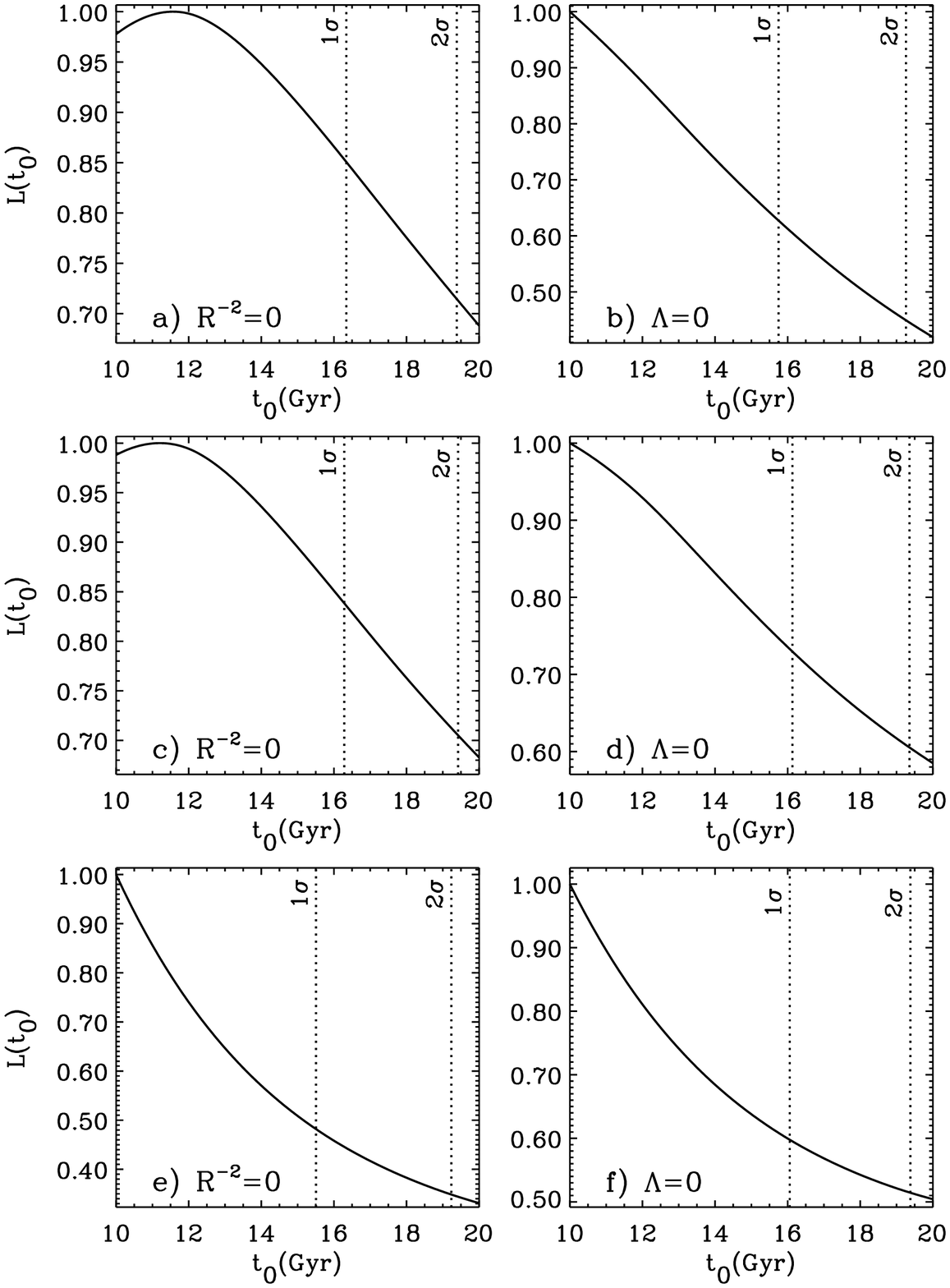}
\end{center}  
\vfill
Figure 7

\clearpage
\begin{center}
  \leavevmode
  \epsfysize=8.0truein
  \epsfbox{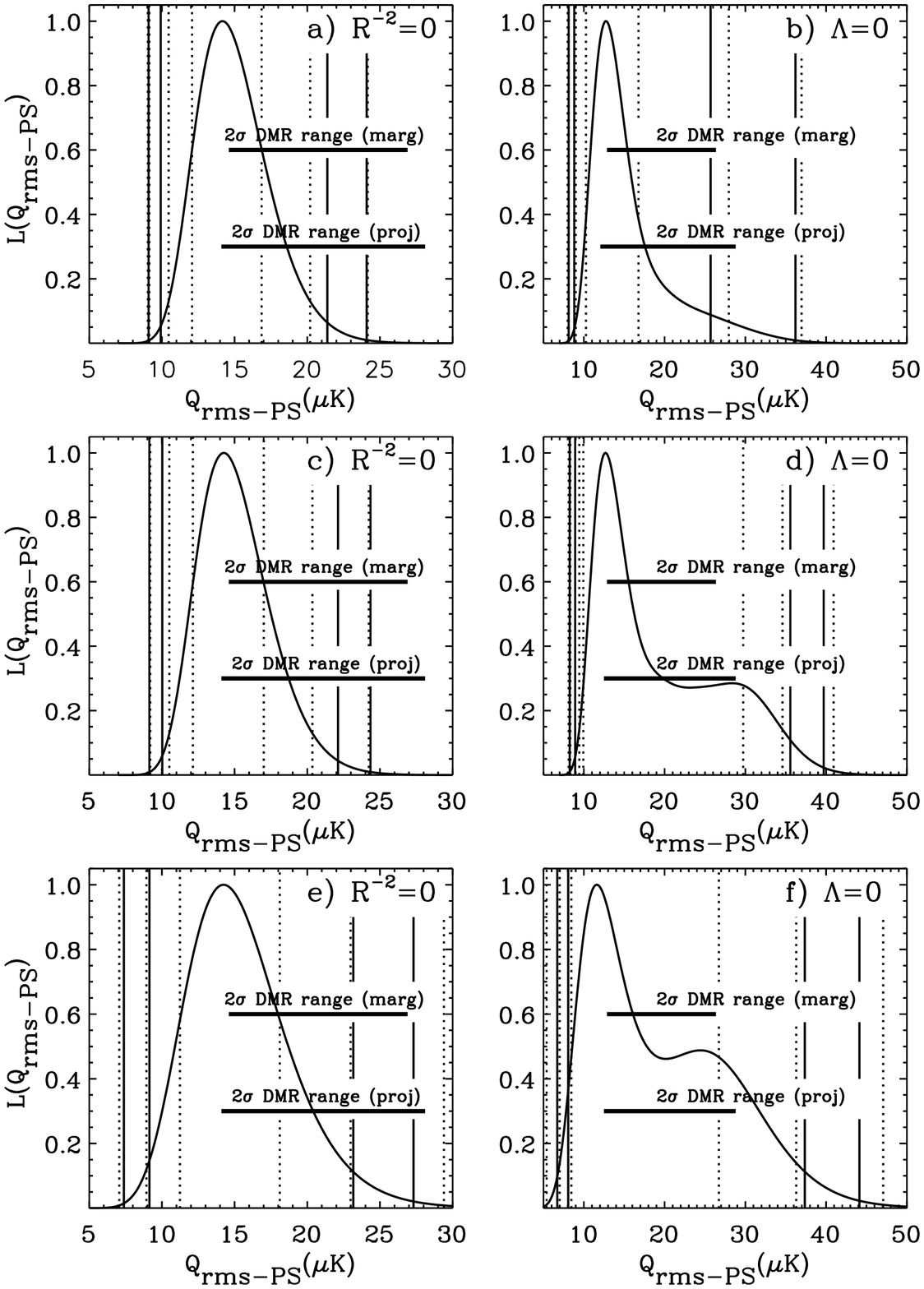}
\end{center}  
\vfill
Figure 8

\end{document}